\documentclass[aps,reprint,pra,superscriptaddress,floatfix,
]{revtex4-1}  %\documentclass[preprint,11pt]{revtex4-1}  %
\usepackage{graphicx,epsfig,epstopdf}
\usepackage[dvipsnames]{xcolor}
\usepackage{amsmath}
\usepackage{amsthm}
\usepackage{subfigure}
\usepackage{mathrsfs}
\usepackage{bm}
\usepackage{graphicx}
\usepackage{times}
\usepackage[utf8]{inputenc}
\usepackage{upgreek}
\newtheoremstyle{query}%
{}{}%space above/below
{\color{blue}}%body style
{}%heading indent
{\sffamily\bfseries}{:}{12pt}%heading style/punctuation/space after
{}% head spec
\theoremstyle{query}
\newtheorem{aq}{Author Query/Comment}

\newcommand{\baq}{\begin{aq}}
\newcommand{\eaq}{\end{aq}}

\usepackage{newunicodechar}
\newunicodechar{ﬁ}{fi}

\begin{document}
\preprint{APS/123-QED}

\title{Experimentally Realizing Efficient Quantum Control with Reinforcement Learning}
\author{Ming-Zhong Ai}
\email{These two authors contributed equally to this work.}
\affiliation{CAS Key Laboratory of Quantum Information, University of Science and Technology of China, Hefei 230026, China}
\affiliation{CAS Center For Excellence in Quantum Information and Quantum Physics,
	University of Science and Technology of China, Hefei 230026, China}

\author{Yongcheng Ding}
\email{These two authors contributed equally to this work.}
\affiliation{International Center of Quantum Artiﬁcial Intelligence for Science and Technology (QuArtist) and\\ Department of Physics, Shanghai University, 200444 Shanghai, China}
\affiliation{Department of Physical Chemistry, University of the Basque Country UPV/EHU, Apartado 644, 48080 Bilbao, Spain}

\author{Yue Ban}
\affiliation{Department of Physical Chemistry, University of the Basque Country UPV/EHU, Apartado 644, 48080 Bilbao, Spain}
\affiliation{School of Materials Science and Engineering, Shanghai University, 200444
	Shanghai, China}

\author{Jos\'e  D. Mart\'in-Guerrero}
\affiliation{IDAL, Electronic Engineering Department, University of Valencia,
	Avgda. Universitat s/n, 46100 Burjassot, Valencia, Spain}

\author{Jorge Casanova}
\affiliation{Department of Physical Chemistry, University of the Basque Country UPV/EHU, Apartado 644, 48080 Bilbao, Spain}
\affiliation{IKERBASQUE, Basque Foundation for Science, Plaza Euskadi 5, 48009 Bilbao, Spain}

\author{Jin-Ming Cui}\email{jmcui@ustc.edu.cn}
\affiliation{CAS Key Laboratory of Quantum Information, University of Science and Technology of China, Hefei 230026, China}
\affiliation{CAS Center For Excellence in Quantum Information and Quantum Physics,
	University of Science and Technology of China, Hefei 230026, China}

\author{Yun-Feng Huang}\email{hyf@ustc.edu.cn}
\affiliation{CAS Key Laboratory of Quantum Information, University of Science and Technology of China, Hefei 230026, China}
\affiliation{CAS Center For Excellence in Quantum Information and Quantum Physics,
	University of Science and Technology of China, Hefei 230026, China}

\author{Xi Chen}\email{xchen@shu.edu.cn}
\affiliation{International Center of Quantum Artiﬁcial Intelligence for Science and Technology (QuArtist) and\\ Department of Physics, Shanghai University, 200444 Shanghai, China}
\affiliation{Department of Physical Chemistry, University of the Basque Country UPV/EHU, Apartado 644, 48080 Bilbao, Spain}

\author{Chuan-Feng Li}\email{cfli@ustc.edu.cn}
\affiliation{CAS Key Laboratory of Quantum Information, University of Science and Technology of China, Hefei 230026, China}
\affiliation{CAS Center For Excellence in Quantum Information and Quantum Physics,
	University of Science and Technology of China, Hefei 230026, China}

\author{Guang-Can Guo}
\affiliation{CAS Key Laboratory of Quantum Information, University of Science and Technology of China, Hefei 230026, China}
\affiliation{CAS Center For Excellence in Quantum Information and Quantum Physics,
	University of Science and Technology of China, Hefei 230026, China}

\date{\today}

\begin{abstract}
Robust and high-precision quantum control is crucial but challenging for scalable quantum computation and quantum information processing. Traditional adiabatic control suffers severe limitations on gate performance imposed by environmentally induced noise because of a quantum system's limited coherence time. In this work, we experimentally demonstrate an alternative approach {to quantum control} based on deep reinforcement learning (DRL) on a trapped $^{171}\mathrm{Yb}^{+}$ ion. In particular, we find that DRL leads to fast and robust {digital quantum operations with running time bounded by shortcuts to adiabaticity} (STA). Besides, we demonstrate that DRL's robustness against both Rabi and detuning errors can be achieved simultaneously without any input from STA. Our experiments reveal a general framework of digital quantum control, leading to a promising enhancement in quantum information processing.

\end{abstract}

\maketitle

\section{INTRODUCTION}
Two-level systems physically realize qubits, which are the basic units of digital quantum computing. In this paradigm, externally controllable parameters should be designed to manipulate the qubits, implementing fast and robust gate operations. Thus, one can construct a universal fault-tolerant quantum computer with physical platforms based on trapped ions and superconducting circuits~\cite{nielsen2010quantum}. In this way, quantum error correction can also be realized physically to reduce the effects of quantum noises and systematic errors. From this perspective, quantum control is bridged to quantum information processing and quantum computing. This connection leads to enormous researches devoted to producing precise quantum control of qubits with driving fields, including adiabatic passages~\cite{kral2007colloquium}, optimized resonant $\pi$ pulses~\cite{remizov2015synchronization}, composite pulses~\cite{brown2004arbitrarily,torosov2011high,rong2015experimental}, pulse-shape engineering~\cite{steffen2007shaped,barnes2012analytically,daems2013robust}, and other optimizations~\cite{glaser2015training,caneva2009optimal,guerin2011optimal,hegerfeldt2013driving,garon2013time,van2017robust,arenz2017roles}. A most straightforward approach to transit less dynamics obeys the adiabatic theorem by tuning the time-dependent parameter sufficiently slow. However, prolonged operation time destructs the quantum information by induced decoherence, affecting information processing efficiency. 

The concept of shortcuts to adiabaticity (STA)~\cite{guery2019shortcuts,torrontegui2013shortcuts} is proposed, which combines the advantages of both adiabatic passages and resonant pulses. It breaks the adiabatic regime by various techniques, including inverse engineering~\cite{chen2010fast}, counter-diabatic driving~\cite{deffner2014classical,an2016shortcuts}, fast-forward scaling~\cite{masuda2010fast,masuda2012acceleration}, which has been well developed over the past decade. Specifically, inverse engineering emanates from the Lewis-Riesenfeld theory, allowing superadiabatic state evolution on dynamical modes with boundary conditions. {In addition, inverse engineering leaves enough freedom to further allow other tasks such as}, e.g., suppressing systematic errors by collaborating with optimal control theory~\cite{daems2013robust,ruschhaupt2012optimally,lu2013fast}, dynamical decoupling {techniques}~\cite{munuera2020robust}, and machine learning {methods}~\cite{zahedinejad2016designing,liu2019plug,ding2020breaking}. However, invariant-based STA requires continuously tunable parameters, limiting the genre of quantum control as analog-only. We consider a more complicated task: designing digital pulses instead of an analog controller with the same output and similar features. {In this manner, we would deliver a framework that can be naturally integrated in current quantum computing paradigms based on the application of several digital quantum gates.} 
%\JCc{[I was looking for a strong sentence to end this part. But remove it if you fell it is not appropriate.]}

We look for the optimal digital pulses design, which is similar to invariant-based STA for realizing robust quantum control. The optimal design is indeed a combinational optimization problem, being equivalent to dynamic programming, which is no longer analytically solvable.  As artificial intelligence approach,  Reinforcement Learning is a well-known tool for system control~\cite{Sutton2018},  and deep learning has been developed for conquering complicated tasks in many areas~\cite{mnih2015human,mnih2013playing,silver2016mastering,silver2017mastering}, later applied in studying physics~\cite{carleo2017solving,nagy2019variational,hartmann2019neural,vicentini2019variational,yoshioka2019constructing,iten2020discovering}. The framework of deep learning can be combined with reinforcement learning, searching control pulses for quantum state preparation~\cite{henson2018approaching,zhang2019does}, gate operation~\cite{an2019deep}, and quantum Szilard engine~\cite{sordal2019deep}. Since recent researches have employed Deep Reinforcement Learning (DRL) for quantum control~\cite{bukov2018reinforcement,porotti2019coherent,niu2019universal,zhang2018automatic,wu2019learning,wang2020deep}, we are inspired to investigate the connection between DRL and STA. An optimistic expectation is that one can extend STA's concept, introducing DRL as a new technique if it learns the features of STA protocols.

In this paper, we present an experimental demonstration of a robust and high-precision quantum control task based on the deep reinforcement learning method on a trapped $^{171}\mathrm{Yb}^{+}$ ion. To be more specific, we train an Agent in a computer through DRL to achieve a single qubit X gate with time prior information bounded by STA. The multi-pulses control sequences produced by the DRL model is more robust than the standard $\pi$ pulse method (interacting with a constant amplitude for a period of time) with constant Rabi frequency %\JCc{[Is this the standard top-hat $\pi$ pulse? If so, it would be good to clarify here to better guide the reader.]} 
in the presence of system noise. Besides, the robustness against both Rabi and detuning errors at the same time by DRL sequences is also verified. To demonstrate the application in the real laboratory noise environment, we examine the DRL models in the Zeeman energy level of the ion, which is sensitive to magnetic field noise. The results show that these DRL models can combat real system noises.  

\begin{figure*}
\includegraphics{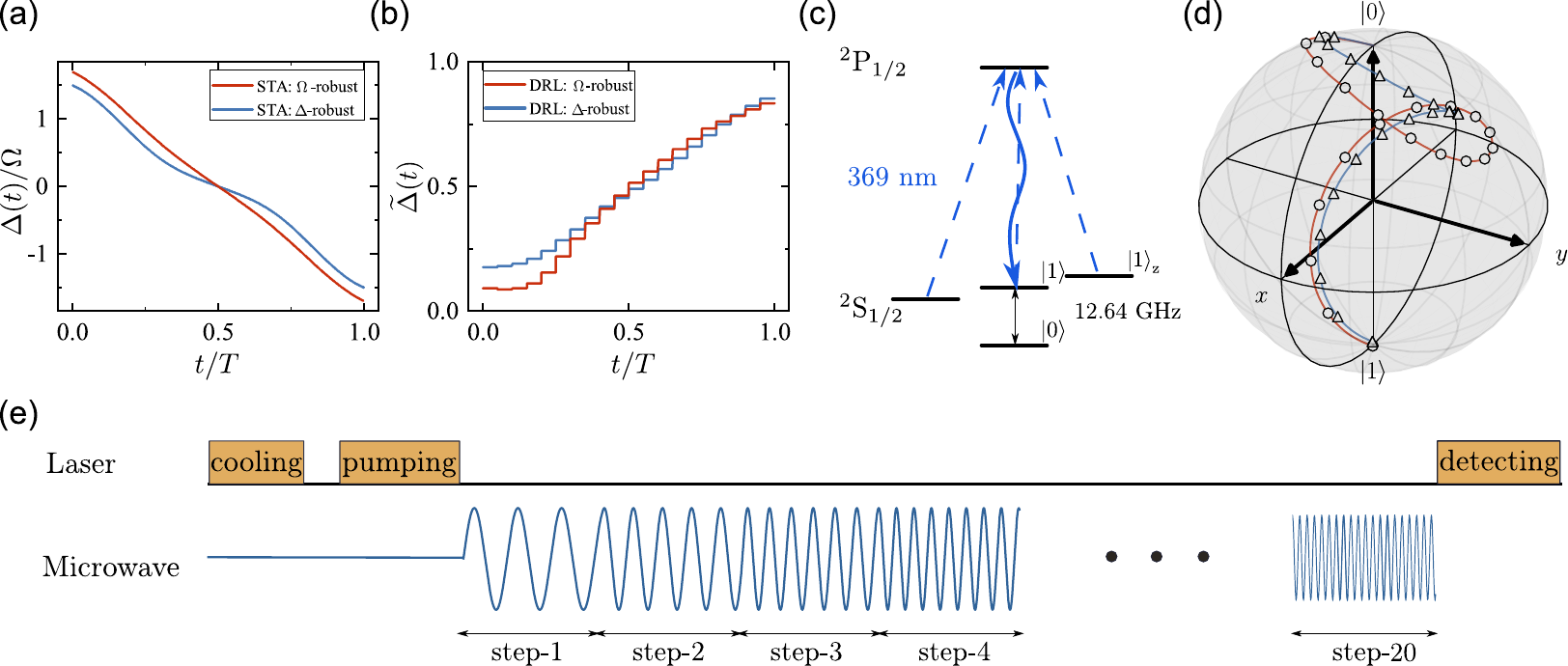}\caption{(color online) Experimental sequences and model wave-forms. (a) Optimized detuning with time under STA method. The time is normalized to $[0,1]$. (b) Optimized detuning with time under DRL method. The time is normalized to $[0,1]$. (c) Energy level of $^{171}\textrm{Yb}^{+}$ ion. (d) Evolution of state in Bloch sphere under the driving of DRL model. Red solid line represents the trajectory optimized for $\Omega$ errors while blue solid line represents the trajectory optimized for $\Delta$ errors. Hollow circle and hollow triangle represent the state at the end of each driving step. (e) Experimental sequences in DRL model. After laser cooling and pumping, the ion is initialized to $|0\rangle$ state. Then a 20-steps microwave which contains DRL driving information is transmitted to the ion. finally a detecting laser is used to detect the probability in $|1\rangle$ state of the ion.}
\end{figure*}

\section{THEORETICAL MODELS}
Consider the coherent manipulation of a single qubit, whose Hamiltonian reads
\begin{equation}
H=\frac{\hbar}{2}\left[\Omega\sigma_x+\Delta(t)\sigma_z\right],
\end{equation}
where the Rabi frequency $\Omega$ is fixed, while the detuning $\Delta(t)$ is time-varying. To achieve a robust qubit flipping from $|0\rangle$ to $|1\rangle$, a standard $\pi$ pulse, which corresponds to the Hamiltonian $\frac{\hbar}{2} \Omega \sigma_{x}$, is convenient and adequate. However, this operation is sensitive to systematic noise and decoherence.

The invariant-based STA suggests that, one can achieve nonadiabatic quantum control of high fidelity and robustness by designed protocols, which satisfy the auxiliary equations derived from Lewis-Riesenfeld (LR) invariant. The LR invariant of a two-level system is constructed by $I(t)=\frac{\hbar}{2}\Omega_0\sum_\pm|\psi_\pm(t)\rangle\langle\psi_\pm(t)|$, where the eigenstates are $|\psi_+(t)\rangle=\left(\cos\frac{\theta}{2}e^{-i\frac{\beta}{2}},\sin\frac{\theta}{2}e^{i\frac{\beta}{2}}\right)^{\text{T}}$ and $|\psi_-(t)\rangle=\left(\sin\frac{\theta}{2}e^{-i\frac{\beta}{2}},-\cos\frac{\theta}{2}e^{i\frac{\beta}{2}}\right)^{\text{T}}$. The dynamics of the Hamiltonian is governed by time-dependent Schr\"odinger's equation, whose solution is in superposition of these eigenstates as $|\Psi(t)\rangle=\sum_\pm c_\pm\exp(i\gamma_\pm)|\psi_\pm(t)\rangle$, with LR phase calculated as
\begin{equation}
\gamma_\pm=\pm\frac{1}{2}\int_0^t\left(\frac{\dot{\theta}\cot\beta}{\sin\theta}\right)dt'.
\end{equation}
According to the condition for invariant $dI(t)/dt=\partial I(t)/\partial t+(1/i\hbar)[I(t),H(t)]=0$, we have {the} auxiliary equations
\begin{eqnarray}
\dot{\theta}&=&-\Omega\sin\beta,\\
\dot{\beta}&=&-\Omega\cot\theta\cos\beta+\Delta(t),
\end{eqnarray}
describing the state evolution along the dynamical modes with angular parameters $\theta$ and $\beta$, which characterize the trajectory on the Bloch sphere. As proposed in Ref.~\cite{ding2020breaking}, the framework can be applied to design robust quantum control, e.g., qubit flipping, against systematic errors with an adequate ansatz of free parameter $a$, {such that}
\begin{equation}
\theta(t)=\frac{\Omega T}{a}\left[as-\frac{\pi^2}{2}(1-s)^2+\frac{\pi^3}{3}(1-s)^3+\cos(\pi s)+A\right],
\end{equation}
where $T=-\pi a/[(2-a-\pi^2/6)\Omega]$, $s=t/T$, and $A=\pi^2/6-1$ determined by boundary conditions $\theta(0)=0,~\dot{\theta}(0)=\Omega,~\ddot{\theta}(0)=0$ and $\theta(T)=\pi,~\dot{\theta}(T)=\Omega,~\ddot{\theta}(T)=0$. Specifically, one can nullify the probability of the first-order transition
\begin{equation}
P=\frac{\hbar^2}{4}\left|\int_0^T\langle\Psi_-(t)|\left(\delta_\Omega\Omega\sigma_x+\delta_\Delta\sigma_z\right)|\Psi_+(t)\rangle\right|^2,
\end{equation}
which yields the condition for error cancellation
\begin{equation}
\left|\int_0^Tdte^{i2\gamma_+(t)}\left(\delta_\Delta\sin\theta-i2\delta_\Omega\dot{\theta}\sin^2\theta\right)\right|=0,
\end{equation}
where systematic errors are characterized by $\Delta(t)\rightarrow\Delta(t)+\delta_\Delta$ and $\Omega\rightarrow\Omega(1+\delta_\Omega)$, resulting in the configuration $a=0.604$ and $0.728$ for eliminating $\Delta$ and $\Omega$-error, respectively. Indeed, smooth detuning pulse $\Delta(t)$ as analog control of single-component is inversely engineered by substituting the ansatz into the following expression
\begin{equation}
\Delta(t)=-\frac{\ddot{\theta}}{\Omega\sqrt{1-\left(\frac{\dot{\theta}}{\Omega}\right)^2}}+\Omega\cot\sqrt{1-\left(\frac{\dot{\theta}}{\Omega}\right)^2}
.
\end{equation}
which is derived from combining auxiliary equations. The wave-forms of $\Delta(t)$ optimized for different systematic errors in STA are shown in fig. 1(a) and the maximum detuning $\Delta_{\max}$ for $\Delta$ and $\Omega$ errors are $1.5\Omega$ and $1.7\Omega$, respectively. Concerning our physical realization in trapped ions, the Rabi frequency $\Omega=(2\pi) 3.3$ kHz is fixed, %{[Why not to write $$ kHz? Later, in page 3, it is used this one]} 
where we calculate the corresponding operation time for robust qubit flipping against $\Delta$ and $\Omega$-errors as $T_\Delta=364$ $\mu$s and $T_\Omega=293$ $\mu$s. 

Since an analog quantum control can be derived from the STA framework, it is more challenging to consider the digital quantum control of Landau-Zener problem. The problem is reformulated to the following expression: how should we manipulate a quantum system for a certain target with a step controller of $N$ intervals within a fixed time? The combinational optimization problem is equivalent to dynamic programming, i.e., a multi-step decision problem whose complexity grows exponentially with step number, allowing an approximation solution by artificial neural networks (ANN) or other universal function approximators; the use of deep ANN architectures with many layers leads to the concept of deep learning,  and this, in turn, to DRL. In the framework of DRL, one assumes that there exists an unknown global optimal policy $\pi$ for a task, which gives an action $\textbf{a}(t_i)$ once observing an arbitrary state $\textbf{s}(t_i)$ at time $t_i$. The state-action relation $\pi(\textbf{s}|\textbf{a})$ is approximated by an Agent ANN, containing propagation of information between layers and nonlinear activation of neurons, whose parameters are tuned by optimizing algorithms for maximizing the accumulated reward. Details about the implementation of deep reinforcement learning can be found in supplementary materials.

In our numerical experiments, the tunable range of detuning $[-\Delta_{\max},\Delta_{\max}]$ is renormalized into $\tilde{\Delta}\in[0,1]$ with $\Delta_{\max}$ being the maximal reachable value of $\Delta(t)$ in STA, which is the output of ANN as the encoded action at time step $t_i$: $\tilde{\Delta}(t_i)=[\Delta(t_i)+\Delta_{\max}]/2\Delta_{\max}$. Information of the two-level system, specifically, the expectation of spin on Z direction $\langle\sigma_z\rangle$, the renormalized detuning $\tilde{\Delta}(t_{i-1})$ that drives the system to the current state, and the system time $i/N$, are fed to the input layer of the ANN. The quantum dynamics are simulated by Liouville-von Neumann equation, which can be generalized to the Lindblad master equation for taking quantum noises into consideration. While network configuration, hyperparameters, and training details are explained in the literature \cite{ding2020breaking}, we introduce the reward functions that we artificially design, which are similar to invariant-based STA that chooses an ansatz for obtaining quantum control. For converging the Agent to robust control of LZ-type, we firstly pre-train the Agent with $r(t_i)=-|\tilde{\Delta}(t_i)-\frac{i-1}{N-1}|$, punishing the deviations from linear growth of detuning, later rewarding a constant if $\langle\sigma_z\rangle>0.997$ at the final time step for fine-tuning under random systematic errors. 

For evaluating the DRL-inspired robust quantum control, we perform two numerical experiments as follows: (\romannumeral1) We set the operation time as $T_\Delta=364$ $\mu$s and $T_\Omega=293$ $\mu$s, being split uniformly by 20 pulses as the only hint from STA. The digital wave-forms output from our DRL model optimized for different systematic errors are shown in fig. 1(b). We emphasize that the STA framework clarifies the upper bound of robustness in Landau-Zener problems, which could be employed for benchmarking the capability of the Agent, as an artificial intelligence approach to digital quantum control with the alike feature. (\romannumeral2) The operation time is arbitrarily set to be $T=300$ $\mu$s for checking if the Agent can explore desired protocols against hybrid systematic errors without any field knowledge of STA. We clarify that DRL is more general for this task since invariant-based STA no longer eliminates the hybrid errors perfectly but on certain proportion of $\delta_\Delta$ and $\delta_\Omega$ instead. All wave-forms used in real experiments are from these two numerical experiment models.

\section{EXPERIMENTAL REALIZATION}
Our experiments are performed on a $^{171}{\rm Yb}^{+}$ ion trapped in a harmonic Paul trap, with the simplified structure being described in detail in supplementary materials. As shown in fig. 1(c), the two level system (TLS) is encoded in the $^{2}{\rm S}_{1/2}$ ground state of the ion, with $\left|0\right\rangle =\left|^{2}{\rm S}_{1/2},F=0,m_{F}=0\right\rangle $ and $\left|1\right\rangle =\left|^{2}{\rm S}_{1/2},F=1,m_{F}=0\right\rangle $. The difference of energy level $\left|0\right\rangle $ and $\left|1\right\rangle $ is about $\omega_{01}=12.6428$ GHz. The microwaves used to drive the TLS are generated through mixing method. More specifically, a microwave around 12.4428 GHz generated from signal generator (Agilent E8257D) is mixed with a 200 MHz microwave signal which is generated from a arbitrary waveform generator (AWG) and is used to modulate the microwave. After a high pass filter (HPF), this signal will be amplified to about 10 W and then transmitted to the ion with a microwave horn \citep{cui2016experimental}. Our trap device is shielded with a 1.5 mm thick single layer Mu-metal \citep{farolfi2019design}, making the final coherence time about 200 ms for $\left|0\right\rangle \leftrightarrow\left|1\right\rangle $ transition, which is characterized by Ramsey experiments.

In each cycle, the experiment takes the following process: after 1 ms Doppler cooling, the state of the ion is initialized to $\left|0\right\rangle $ state through 20 $\mu$s optical pumping with $99.5\%$ fidelity. The wave-form output from DRL model is transformed into driving microwave through modulating the detuning, which is shown in fig. 1(e). Then the driving microwave is transmitted to the ion to drive the TLS. finally, a NA (numerical aperture) = 0.4 objective is used for state dependent fluorescence detection to determine the probability in state $\left|1\right\rangle $. In all of our experiments we set the Rabi frequency to $\Omega=(2\pi)$ 3.3 kHz, that is to say, the corresponding $2\pi$ time is about 300 $\mu$s. 

\begin{figure}
\includegraphics{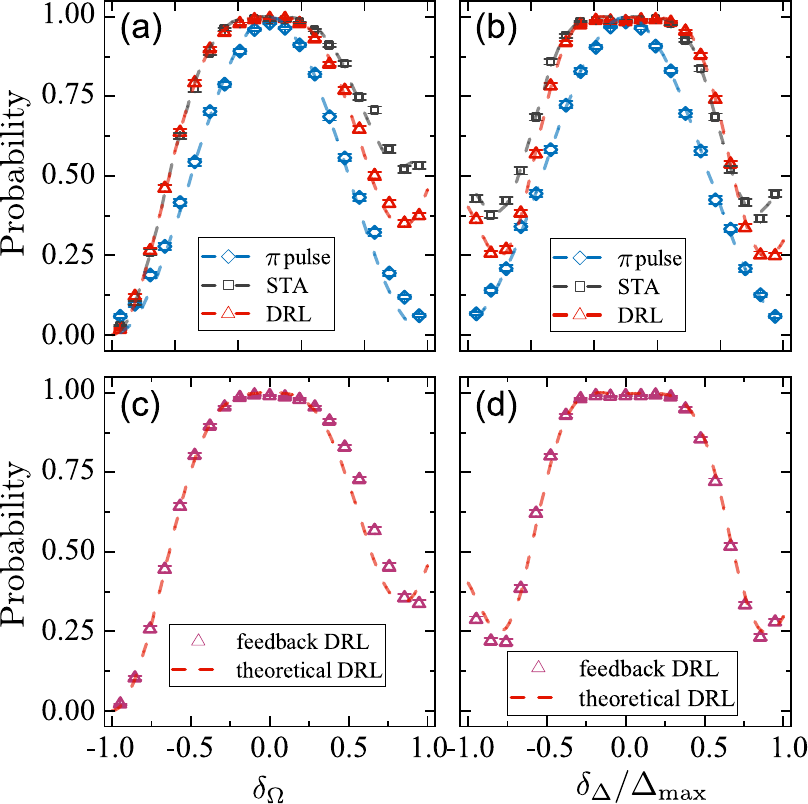}\caption{(color online) Noise robustness comparison of $\pi$ pulse, STA and DRL methods in single-qubit X gate task. (a) and (b) The performance of three control methods under different $\Omega$ and $\Delta$ errors respectively. The DRL method is as robust as STA method in most cases, except in big $\Omega$ and $\Delta$ errors. But they are all more robust than $\pi$ pulse in both kinds of errors. (c) and (d) The performance of feedback DRL. The feedback DRL agrees well with theoretical DRL in both $\Omega$ and $\Delta$ errors, which indicates our DRL model is robust to the disturbance of control pulses. The error bars indicate the standard deviation, and each data point is averaged over 2000 realizations.}
\end{figure}

To verify the robustness of the DRL control method against systematic errors, we compare the performance of STA, DRL, and standard $\pi$ pulse method in the single-qubit X gate task under different $\Delta$ and $\Omega$ errors. The DRL models are pre-trained according to the time preliminary information provided by STA methods optimized in $\Delta$ and $\Omega$ errors, respectively. The state evolution under STA driving in Bloch sphere is shown in fig. 1(d). As shown in fig. 2 (a) and 2 (b), the DRL method performs as well as STA in most cases, in addition to the case that $\Omega$ error or $\Delta$ error is too large. Meanwhile, they are all more robust than the $\pi$ pulse method under system errors. To further explore our DRL model's robustness, we also perform a feedback DRL experiment. In this experiment, 19 cycles are carried out. In cycle $n$, where $n$ belongs to [1,19], we measure the experimental result after $n$ control pulses, and feedback this result to the DRL model to obtain the next control pulses. After 19 cycles, we get the final 20 control pulses, and these pulses are only a little different from theoretical DRL pulses. As shown in fig. 2 (c) and 2 (d), the experimental feedback DRL results agree well with theoretical DRL, which means that our DRL model is robust to the disturbance of control pulses.

\begin{figure}
\includegraphics{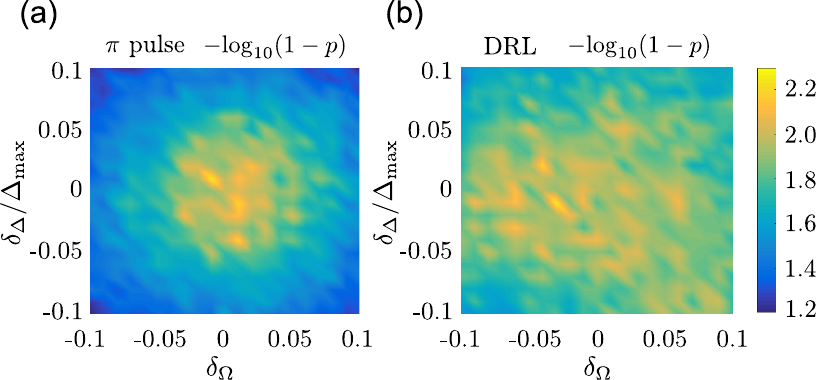}\caption{(color online) The performance of $\pi$ pulse and DRL model under hybrid errors. The $\pi$ pulse method performs a little better than DRL in the case of almost no errors, because the pulses of DRL are more complex that it is easy to accumulate operation errors. However, with the increase of hybrid errors, the performance of DRL model is much better than $\pi$ pulse control.}
\end{figure}

Then we examine the DRL model under $\Delta$ and $\Omega$ hybrid errors. It is worthwhile to mention that we set operation time and tunable range of detuning without any knowledge from STA when pre-training the DRL model locally. The performance of $\pi$ pulse and DRL method under hybrid errors is shown in fig. 3, in which the probabilities are taken logarithm to better distinguish the difference between these two methods The DRL method is more likely to accumulate errors than $\pi$ pulse due to the multi pulses driving operation on the one hand, on the other hand we just stop our training once $\langle \sigma_{z} \rangle > 0.997$, which can be further improved theoretically. As we can expect, the $\pi$ pulse method performs a little better than DRL in the case of almost no errors. Nevertheless, with the increase of hybrid errors, the DRL method's performance is much better than $\pi$ pulse in most cases, which is essential in precise quantum manipulation.

Besides, we also examine the DRL model in the Zeeman energy level of the ion with $\left|1\right\rangle _{\textrm{z}} =\left|^{2}{\rm S}_{1/2},F=1,m_{F}=1\right\rangle $. The Zeeman energy level is first-order sensitive to the disturbance of the magnetic field, which could induce the realistic laboratory noise into TLS, and the corresponding coherence time is about 0.35 ms for $\left|0\right\rangle \leftrightarrow\left|1\right\rangle _{\textrm{z}}$ transition. The experimental results demonstrate that DRL's performance is a little worse than theoretical expectation both in $\Omega$ and $\Delta$ errors due to extra decoherence, which is shown in fig. 4 (a) and 4 (b). We also compare the performance of $\pi$ pulse and DRL method in the single-qubit X gate under only the magnetic field noise with different Rabi time and different number of $\pi$ flips. As shown in fig. 4 (c) and 4 (d), the final probability decreases rapidly with the Rabi time and number of $\pi$ flips in $\pi$ pulse method owing to inevitable decoherence. However, the DRL method is more robust with the increase of Rabi time and number of $\pi$ flips, which is important in noisy quantum information processing. 

\begin{figure}
\includegraphics{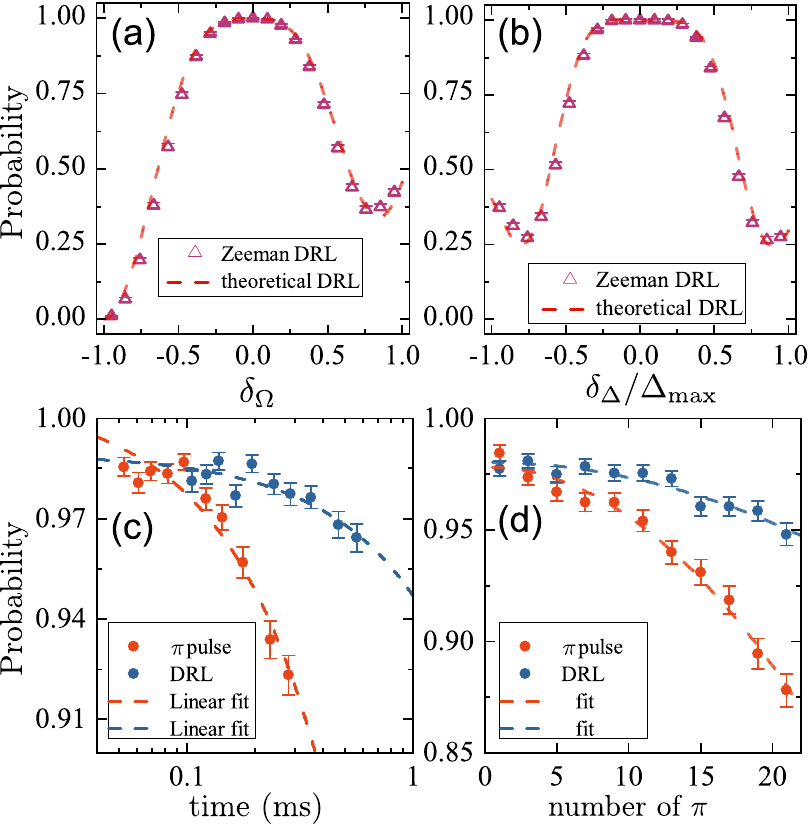}\caption{(color online) Noise-resilient feature of DRL method in Zeeman energy level. (a) and (b) The performance of DRL in Zeeman energy level. The DRL performs a little worse than theoretical values both in $\Omega$ and $\Delta$ errors due to disturbance of laboratory noise. (c) The comparison of $\pi$ pulse and DRL under different Rabi time. With the increase of Rabi time, the performance of $\pi$ pulse decreases rapidly while the DRL is more robust against decoherence. (d) The comparison of $\pi$ pulse and DRL under different number of $\pi$ flips. With the increase of $\pi$ flips, the performance of $\pi$ pulse decreases rapidly while the DRL is more robust against decoherence. The error bars indicate the standard deviation, and each data point is averaged over 2000 realizations}
\end{figure}

\section{CONCLUSION}
In summary, we experimentally demonstrate a robust quantum control task based on deep reinforcement learning. The DRL model's multi-pulse control sequences are more robust than $\pi$ pulse in the presence of systematic errors. We also verify that robustness against both Rabi and detuning errors simultaneously can be achieved by DRL without any input from STA. In addition, we confirm that these DRL models can be significant in the real laboratory environment, which will lead to a promising enhancement in quantum information processing. 
\begin{acknowledgments}
%\bigskip
This work was supported by the National Key Research and Development Program of China (Nos. 2017YFA0304100, 2016YFA0302700), the National Natural Science Foundation of China (Nos. 11874343, 61327901, 11774335, 11474270, 11734015, 11874343), Key Research Program of Frontier Sciences, CAS (No. QYZDY-SSW-SLH003), the Fundamental Research Funds for the Central Universities (Nos. WK2470000026, WK2470000018), An-hui Initiative in Quantum Information Technologies (AHY020100, AHY070000), the National Program for Support of Topnotch Young Professionals (Grant No. BB2470000005). The theoretical part of the work is also partially supported from NSFC (12075145), STCSM (2019SHZDZX01-ZX04, 18010500400 and 18ZR1415500), Program for Eastern Scholar, HiQ funding for developing STA (YBN2019115204), QMiCS (820505) and OpenSuperQ (820363) of the EU Flagship on Quantum Technologies, Spanish Government PGC2018-095113-B-I00 (MCIU/AEI/FEDER, UE), Basque Government IT986-16, EU FET Open Grant Quromorphic (828826) as well as EPIQUS (899368). X. C. acknowledges Ram\'on y Cajal program (RYC-2017-22482). {J. C. acknowledges the Ram\'on y Cajal program (RYC2018-025197-I) and the EUR2020-112117 project of the Spanish MICINN, as well as support from the UPV/EHU through the grant EHUrOPE.}
\end{acknowledgments}

\pagebreak
\widetext
\begin{center}
	\textbf{ \large Supplemental Material: \\ Experimentally Realizing Efficient Quantum Control with Reinforcement Learning}
\end{center}

%%%%%%%%%% Merge with supplemental materials %%%%%%%%%%
%%%%%%%%%% Prefix a "S" to all equations, figures, tables and reset the counter %%%%%%%%%%
\setcounter{equation}{0} \setcounter{figure}{0} \setcounter{table}{0}
\global\long\def\thefigure{S\arabic{figure}}
\global\long\def\bibnumfmt#1{[S#1]}
\global\long\def\citenumfont#1{S#1}
\setcounter{section}{0}

\section{EXPERIMENTAL PLATFORM AND WAVEFORM OF THE DRIVING MICROWAVE}

The type of platform used in our experiments is needle trap. As shown in Fig. S1, the needle trap consists of 6 needles. Two opposite needles are connected to radio frequency (RF) potential to trap the ion and the others are connected to direct current (DC) potential to fine tuning the position of ion. The size of the needle trap depends mainly on the distance between the two needles tips near the trap center, which is set to 180 $\mu$m in our experiment. The trap is installed in an ultrahigh vacuum below $10^{-11}$ torr, and a helical resonator provides the RF signal with frequency 24 MHz and amplitude of 180 V to the trap. Ion fluorescence is collected by an objective lens with 0.4 numerical aperture, and detected by a photo-multiplier tube (PMT). The total fluorescence detection efficiency is about 2$\%$. 

We generate required waveform of the microwave field through setting the waveform of AWG for modulation. The carrier microwave $B_{c}(t)=A_{1}\mathrm{sin}(\omega_{c} t)$, where $A_{1}$ is amplitude and $f_{c}=\omega_{c}/2\pi=12.4$ GHz is the frequency. The waveform generated by AWG for modulation is $I(t)=A_{2}\mathrm{sin}(\phi(t))$. After mixing, the microwave field will be $B(t)=\frac{A_{1}A_{2}}{2}(\mathrm{sin}(\omega_{c}t + \phi(t)) + \mathrm{sin}(\omega_{c}t - \phi(t)))$, where the phase function $\phi(t)$ can be expressed in a piece-wise function for the microwave composed of 20 steps in our DRL experiments. With the qubit resonance frequency $f_{0}=\omega_{0}/2\pi = 12.6$ GHz, we filter out the low frequency components of the microwave through a high pass filter. In our experiments, we only adjust $\Delta(t)$ with discrete steps by changing phase $\phi(t)$ as follows:

\begin{equation}
\phi(t)=\begin{cases}
(\omega_{0}-\omega_{c})t+\Delta_{1}t, & (0,t_{1})\\
(\omega_{0}-\omega_{c})t+\Delta_{2}t+\phi_{1}, & (0,t_{2}-t_{1})\\
(\omega_{0}-\omega_{c})t+\Delta_{3}t+\phi_{2}, & (0,t_{3}-t_{2})\\
\cdots\\
(\omega_{0}-\omega_{c})t+\Delta_{20}t+\phi_{19}, & (0,t_{20}-t_{19})
\end{cases}
\end{equation}
where $\Delta_{n}(n\in[1,20])$ is the step-wise detuning and $\phi_{1}=(\omega_{0}-\omega_{c})t_{1} + \Delta_{1}t_{1}$, $\phi_{2}=(\omega_{0}-\omega_{c})(t_{2}-t_{1}) + \Delta_{2}(t_{2}-t_{1}) + \phi_{1}$ and so on.

\begin{figure}[!h]
\includegraphics{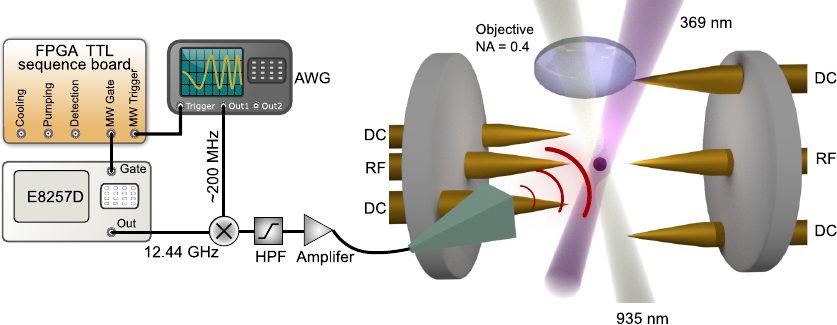}\caption{\label{fig:} (color online)  Experimental setup. A single $^{171}\rm{Yb}^{+}$ ion is trapped in center of the needle trap. Two 369 nm and 935 nm lasers are used to cooling the ion and 369 nm laser is also used to detect the state of ion. The microwave used to drive the ion is generated through mixing method. The whole experimental sequences are controlled by a TTL sequences board based on Field Programmable Gate Array (FPGA).}
\end{figure}

\section{QUBIT STATE PREPARATION AND MEASUREMENT}

In ion trap experiments, the state preparation and measurement cannot be perfect and there will always be some limitations. We characterize these errors as follows. The ion is prepared in $|0\rangle$ state through optical pumping and ideally, no photon should be detected as the ion is in dark state. However, due to the dark counts of the photon detector as well as photons scattered from the environment, we will collect some photons sometimes. Then we apply a $\pi$ pulse to flip the $|0\rangle$ state to $|1\rangle$ state and detect the fluorescence. Because the collection efficiency problem, no photon will be collected sometimes. The histograms of dark and bright state is shown in Fig. S2. The threshold is selected as 2 in our experiments. When the photon number is $>2$, the qubit is identified as bright state and the probability of being mistaken as dark state is $\epsilon_{D}$. By contrary, the probability of being mistaken as bright state when photon number is $\leq2$ is $\epsilon_{B}$. The total error can be taken as $\epsilon=(\epsilon_{B} + \epsilon_{D})/2$.

\begin{figure}
\includegraphics{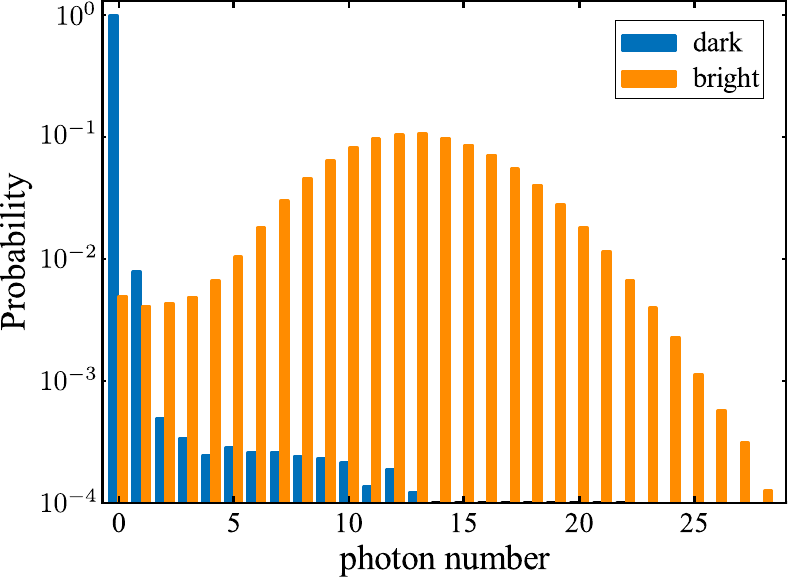}\caption{\label{fig:} (color online) Histogram for photon counts in state preparation and detection
experiments. The distribution of photon counts is shown when the qubit
state is prepared in $|0\rangle$ (dark state) and $|1\rangle$ (bright state).}
\end{figure}

\section{IMPLEMENTATION OF DEEP REINFORCEMENT LEARNING}
By combining reinforcement learning and deep learning, deep reinforcement learning (DRL) aims to solve decision-making problems, allowing a computational agent to make decisions from input data by trial. A mathematical model called Markov decision process describes the problem, where an agent at every time step $t$ observes a state $s_t$, takes an action $a_t$, receives a reward $r_t$ and transits to the state at the next time step $s_{t+1}$ according to the dynamics of the environment $P(s_{t+1}|s_{t},a_t)$. The agent's goal is to learn a policy $\pi(a|s)$ that maximizes the total reward $\sum_i\gamma^ir_i$, with $\gamma$ being the discount rate and $r_i$ being the scalar rewards. DRL employs an artificial neural network (ANN) as a general function approximator for the policy $\pi(a|s)$, leading to specialized algorithms for obtaining an optimal approximation. The simplified model framework can be found in Fig. S3.

\begin{figure}
\includegraphics{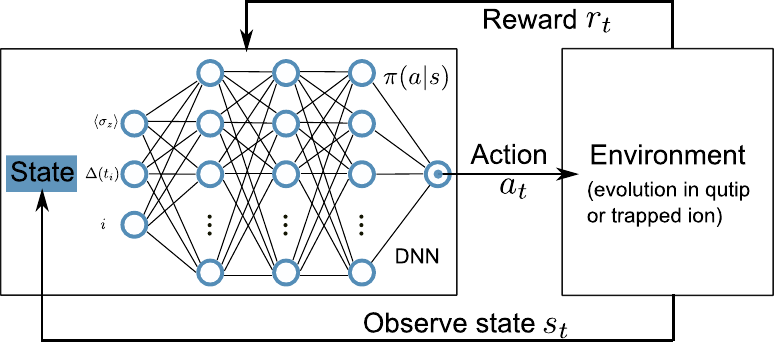}\caption{\label{fig:} (color online) DRL framework for quantum control with qubit of one time step in training. The agent (DNN), consists of three hidden layers, observes a state from environment. After propagation between layers and nonlinear activations of the DNN nodes, output layer gives an action $a_{t}$. The environment rewards $r_{t}$ enable the agent to learn how to achieve the goal.}
\end{figure}

Here we use Proximal Policy Optimization (PPO), which performs comparably state-of-the-art, remaining simplicity for implementation and tuning. It is worthwhile to mention that it is also the default RL algorithm at OpenAI. As an on-policy algorithm, PPO attempts to evaluate and improve the behavior policy that is used to make decisions. Its objective function is
\begin{equation}
L_{\text{clip}}(\theta)=\hat{E}_t\left\{\min\left[r_t(\theta)\hat{A}_t,\text{clip}\left(r_t(\theta),1+\epsilon,1-\epsilon\right)\hat{A}_t\right]\right\},
\end{equation},
where $\theta$ is the policy parameter (the set that contains all weights and biases), $\hat{E}_t$ is the expectation over time steps, $r_t$ is the ratio of the probability under the new and old policies, $\hat{A}_t$ is the estimated advantage, and $\epsilon$ is the hyperparameter for bounding the clipping range. There is also a variant of PPO based on an adaptive Kullback-Leiber penalty, which controls the change of $\pi(a|s)$ at each iteration. A detailed explanation of PPO, as well as its pseudocodes, are already clearly presented in the original paper~\cite{schulman2017proximal}. Although there are arguments about the origin of performance enhancement from Trust Region Policy Optimization~\cite{schulman2015trust} (whether it is from the clipping or code-level tricks~\cite{engstrom2019implementation}), we reckon these topics, including if PPO-like algorithms can be further optimized, go beyond the scope of this work. Thus, we implement a minimal PPO for our quantum control task.

We use an open-source Python library, TensorForce (version 0.5.2)~\cite{schaarschmidt2017tensorforce}, for a quick implementation. The library is based on TensorFlow, a well-known framework for deep learning with GPU acceleration. The two-level system's quantum dynamics in our training environment are numerically simulated by QuTiP (version 4.4.1)~\cite{johansson2012qutip}. We set a batch size of 16, and the learning rate is 1e-4 for both pre-training and fine-tuning. The ANN contains three hidden layers, where each of the layers consists of 32 fully-connected neurons activated by ReLU. Other hyperparameters and settings are the default configuration of the PPO Agent provided by TensorForce. Another evaluation environment can interact with the trapped ion system for verifying quantum control with feedback. Codes are compatible with both CPU and GPU versions of TensorFlow 1.13.1., which are available from the corresponding authors upon reasonable request.

\end{document}